# Monolayered MoSe$_2$: A candidate for room temperature polaritonics

N. Lundt[1], A. Maryński[2], E. Cherotchenko[3], A. Pant[4], X. Fan[4], G. Sęk[2], S. Tongay[4], A. V. Kavokin[3,5], S. Höfling[1,6], C. Schneider[1]

1) Technische Physik, University of Würzburg, Am Hubland, D-97074 Würzburg, Germany
2) Laboratory for Optical Spectroscopy of Nanostructures, Division of Experimental Physics, Faculty of Fundamental Problems of Technology, Wroclaw University of Science and Technology, Wybrzeże Wyspiańskiego 27, 50-370 Wrocław, Poland
3) Physics and Astronomy School, University of Southampton, Highfield, Southampton, SO171BJ, UK
4) School for Engineering of Matter, Transport, and Energy, Arizona State University, Tempe, Arizona 85287, United States
5) SPIN-CNR, Viale del Politecnico 1, I-00133 Rome, Italy
6) SUPA, School of Physics and Astronomy, University of St. Andrews, St. Andrews KY 16 9SS, United Kingdom

Abstract

Monolayered MoSe$_2$ is a promising new material to investigate advanced light-matter coupling as it hosts stable and robust excitons with comparably narrow optical resonances. In this work, we investigate the evolution of the lowest lying excitonic transition, the so-called A-valley exciton, with temperature. We find a strong, phonon-induced temperature broadening of the resonance, and more importantly, a reduction of the oscillator strength for increased temperatures. Based on these experimentally extracted, temperature dependent parameters, we apply a coupled oscillator model to elucidate the possibility to observe the strong coupling regime between the A-exciton and a microcavity resonance in three prototypical photonic architectures with varying mode volumes. We find that the formation of exciton-polaritons up to ambient conditions in short, monolithic dielectric and Tamm-based structures seems feasible. In contrast, a temperature-induced transition into the weak coupling regime can be expected for structures with extended effective cavity length. Based on these findings, we calculate and draw the phase diagram of polariton Bosonic condensation in a MoSe$_2$ cavity.

Introduction

Inspired by the discovery of graphene, the field of two-dimensional (2D) materials has rapidly extended to a larger variety of atomically thin materials. Within this field, the group of transition metal dichalcogenides (TMDCs) has attracted great attention due to their unique physical properties[1]. In contrast to graphene, monolayers of materials such as $MoS_2$, $WS_2$, $MoSe_2$, $WSe_2$ and $MoTe_2$ exhibit a direct bandgap[2]. This property allows their application as optoelectronic devices[1]. While devices such as light emitting diodes[3,4], solar cells[5], ultra-fast photodetectors[6] and single-photon emitters[7–11] have been demonstrated, TMDCs are also very promising materials to study light matter interactions on a fundamental basis. The coupled spin and valley physics lead to effects such as the valley hall effect and the valley coupling to the optical helicity[12,13]. Furthermore, electronic and optical properties are governed by strongly bound excitons with binding energies up to 0.55 eV[14]. In principle, these very distinct and robust excitonic features promise pronounced light-matter interaction and the observation of strong light matter coupling phenomena, excitonic and polariton condensation, or laser emission based on excitonic gain material when the layers are embedded in a suitable microresonator geometry[15]. In fact, TMDC monolayers have been successfully coupled to photonic crystals[16,17], plasmonic structures[18–20] and micro-cavities[21–25]. The regime of strong coupling, thus far, could be demonstrated most convincingly at cryogenic temperatures[25], while normal mode crossings at room temperature have been reported based on a $MoS_2$ layer in a monolithic Bragg cavity with significantly smaller visibility of the characteristic Rabi doublet[21]. Although, even high quality $MoS_2$ monolayers still suffer from strong, defect-induced emission broadenings, temperature-induced linewidth broadening and intensity quenching are additional dominant limiting factors in high temperature cavity QED (quantum electrodynamics) experiments.

Here, we discuss temperature dependent reflectivity measurements on a $MoSe_2$ monolayer in order to quantify the relevant temperature dependent parameters of the exciton resonance. Namely, we study the linewidth and area of the absorption resonance, a relative measure for the oscillator strength. These two are crucial parameters determining the coupling with an optical mode. Based on these results, we analyse the hypothetical temperature dependence of the normal mode coupling and its visibility parameter for a variety of a commonly used photonic structures, including open-cavity approaches, monolithic Bragg structures and Tamm-plasmon based devices. Calculations are based on the coupled oscillator model as well as on the numerical transfer matrix (TM) simulations. We finally present the polariton condensation phase diagram for $MoSe_2$ calculated assuming the thermal equilibrium Bose-Einstein condensation in a finite size system, to elucidate the possibility of observing the polariton condensation at ambient conditions[26].

Methods

Monolayer (ML) $MoSe_2$ layers were deposited onto 285 nm thermal oxide on Si wafers via conventional exfoliation from bulk $MoSe_2$ crystals. The $SiO_2$ thickness was chosen to be 285 nm to improve the monolayer contrast. Exfoliated MLs were characterized using Raman and photoluminescence spectroscopy, and their thickness was determined via atomic force microscopy measurements. Micro-reflectivity spectra were taken under white-light illumination of a tungsten halogen lamp (250W, 30 μm pinhole, 10 μm illumination spot size). High spatial resolution was obtained by using an infinity corrected 20 times magnifying long working distance microscope objective with numerical aperture of 0.4. The signal was analysed by a 30 cm monochromator combined with a Si-based CCD. The integration time was 10 seconds per spectrum and 10 spectra were averaged to improve the signal to noise ratio. The light source and setup cover a reliable spectral range from 1.5 to 2.2 eV. Following the convention of references[27,28], the reflectance contrast ΔR/R was obtained according to ΔR/R = ($R_{Sample}$ – $R_{Substrate}$)/$R_{Substrate}$ whereas $R_{Sample}$ is the reflectivity of the monolayer on the substrate and $R_{Substrate}$ is the reflectivity of the uncovered substrate. The excitonic absorption manifests as Gaussian shaped signals in the reflectance contrast spectra. In order to deduce the energy, linewidth and amplitude of the absorption resonances, a background subtraction and fitting process is required. To ensure an appropriate background subtraction, transfer matrix calculations for the reflectivity background without excitonic absorption were carried out (see supplementary S1 for further details and the corresponding error analysis). Even though the acquired amplitude does not provide an absolute absorption value, the product of linewidth and amplitude is a quantity proportional to the exciton oscillator strength[29,30].

Theory:

In order to calculate the Rabi splitting evolution with temperature ℏΩ(T) we used two approaches. First, the following equation, extracted from a coupled oscillator approach was used to account for a temperature-induced quenching of the Rabi-splitting via broadening of the excitonic oscillator [31]:

$$\hbar\Omega(T) = \sqrt{V(T)^2 - \left(\frac{\Delta E_x(T) - \Delta E_c}{2}\right)^2} \quad (1)$$

Here, V(T) is the coupling strength, $\Delta E_x(T)$ is the exciton linewidth and $\Delta E_c$ is the cavity linewidth. V(T) is a function of the oscillator strength f(T), the effective cavity length $L_{eff}$ and the effective number of individual monolayers in the cavity $n_{eff}$:

$$V(T) \sim \sqrt{\frac{f(T) * n_{eff}}{L_{eff}}} \quad (2)$$

The initial values (T = 4K) for V (36 meV) and $\Delta E_c$ (1.6 meV) were taken from reference[25]. Then, ℏΩ(T) was calculated for higher temperatures using the measured, relative values for $f(T)$ and $\Delta E_x(T)$. In addition, the visibility parameter ϑ was calculated according to:

$$\vartheta(T) = \frac{\frac{V(T)}{4}}{\Delta E_x(T) + \Delta E_c} \quad (3)$$

A $\vartheta(T)$ value above 0.25 indicates that the strong coupling regime can be distinctively observed in transmission, reflectivity or PL spectra[32].

Secondly, TM calculations were conducted for the reflectivity of a MoSe$_2$ monolayer, hypothetically integrated into the open cavity design described in reference[25]. The dielectric function $\varepsilon(\omega)$ of the MoSe$_2$ monolayer was modelled as a Lorentz oscillator:

$$\varepsilon(\omega) = \varepsilon_b + \frac{f(T)}{\omega_0^2(T) - \omega^2 - i\Delta E_x(T)\omega} \quad (4)$$

Here, $\varepsilon_b$ is the background dielectric function and $\hbar\omega_0$ is the exciton energy. $\varepsilon_b$ was taken from reference[33] ($\varepsilon_b = 26$) and the initial value for $f(T)$ was adjusted to 0.4 to match the splitting calculated according to equation 1. Linewidth and oscillator strength of $\varepsilon(\omega)$ were adjusted for each temperature according to the reflectivity results. The complex refractive index $\tilde{n}(\omega) = n + ik$ was derived from $\tilde{n}(\omega) = \sqrt{\varepsilon(\omega)}$ and used for the TM calculations. Finally, the reflectivity spectrum for each temperature is simulated with the respective refractive indices assuming a monolayer thickness of 0.65 nm[34]. The splitting is deduced from the spectra and correlated with temperature.

In order to compare the open cavity design with other photonic architectures, the same dielectric functions were taken for additional TM calculations. All three considered photonic architectures are illustrated in figure 1. We consider a fully monolithic cavity consisting of two dielectric mirrors[21] or a Tamm plasmon structure referring to a design described in reference[35]. The monolithic cavity mirrors consist each of eight TiO$_2$/SiO$_2$ layer pairs with $\frac{\lambda}{4n}$ thickness. The MoSe$_2$ monolayer is embedded between two $\frac{\lambda}{4n}$ SiO$_2$ layers, whose thicknesses were adjusted (from 129 nm to 128 nm) to tune the cavity mode in resonance with the exciton energy. The structure design that supports Tamm-plasmon modes consists of the identical bottom dielectric mirror, followed by a SiO$_2$/MoSe$_2$ ML/SiO$_2$ core and a 50 nm layer of gold on the top. Here again, the SiO$_2$ layer thicknesses were adjusted to 114 nm to ensure spectral resonance conditions. The resulting splitting was used to calculate the coupling strength V at 4K according to equation 1. Taking $f(T)$ into consideration, the visibility parameter evolutions $\vartheta(T)$ for the alternative cavity designs were calculated as well.

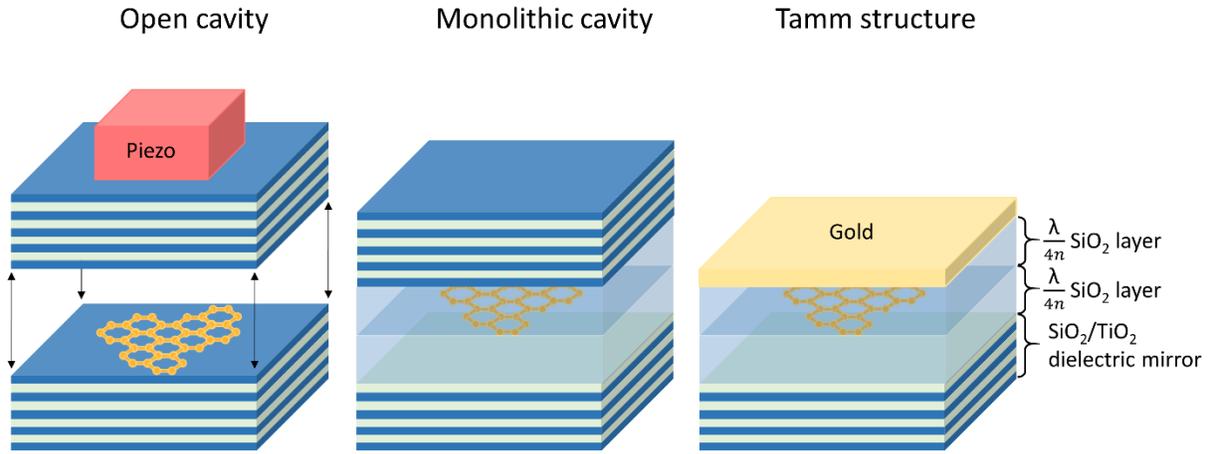

**Figure 1: Schematic illustration of the open cavity design, a fully monolithic cavity and a Tamm plasmon structure.**

In order to check if the strong coupling regime could also lead to polariton Bose-Einstein condensation, we followed the approach in reference[36] to calculate a polariton phase diagram. The phase diagram provides an estimate on the critical polariton density N$_c$ required for polariton condensation at a given temperature T$_c$.

Here we consider a finite system of the lateral size $L$. The particle density is given by:

$$N(T,L,\mu) = \frac{N_0}{L^2} + \frac{1}{L^2}\sum_{\mathbf{k}, k > \frac{2\pi}{L}} \frac{1}{\exp\left(\frac{E(\mathbf{k})-\mu}{k_B T}\right)-1} \qquad (6)$$

where $N_0$ is the population of the ground state, $E(\mathbf{k})$ is the polariton kinetic energy, $\mu$ is the chemical potential, and $k_B$ is the Boltzmann constant.

Defining $N_c$ as the maximum number of particles that can be accommodated in all states but the ground state, one can write:

$$N_c(L,T) = \frac{1}{L^2}\sum_{\mathbf{k}, k > \frac{2\pi}{L}} \frac{1}{\exp\left(\frac{E(\mathbf{k})}{k_B T}\right)-1} \qquad (7)$$

Here $\mu$ is set to be zero that allows to put bosons into the ground state without limitation, while the concentration of polaritons in upper states is constant and equals to $N_c(L,T)$. The condensate density is thus equal to $N_0 = N - N_c$. The upper limit for $N_c$ is assumed to be the Mott density, which is calculated by

$$N_{Mott} = \frac{A}{\pi a_B^2} \qquad (8)$$

where A is the reference area of 1 cm².

## Results and discussion

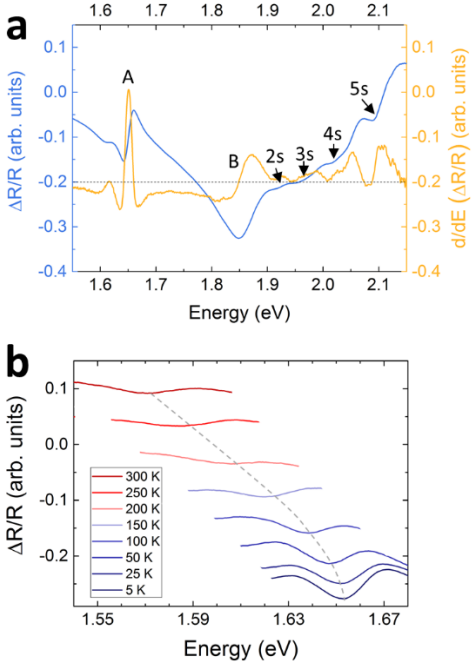

Figure 2: Reflectivity spectra of a MoSe$_2$ monolayer: (a) reflection contrast spectrum (blue) and its derivative (yellow). (b) reflection contrast spectra around the A exciton at various temperature between 5K and 300K.

The micro-reflectivity spectra and their temperature evolution are the experimental basis for the following parameter deduction and calculations. Figure 2a presents a typical reflectance contrast spectrum of a MoSe$_2$ monolayer compared with its derivative for better feature identification. While the A exciton resonance can be clearly identified at 1.653 eV, the B exciton peak at 1.849 eV is quite broadened. This can be explained by the minimum in the background reflectance contrast that is observed at around the same energy. In addition, the absorption peak of the B exciton has in fact been observed to be broader than for the A exciton[37]. Moreover, there are four features above the B exciton which are assigned to higher exciton states and they are therefore labeled 2s, 3s, 4s and 5s. Their energy spacing suggests a non-Rydberg behavior as the larger spatial extension of the higher states compared to the 1s state makes them more sensible to the dielectric screening by the substrate[38]. This effect is well known for WS$_2$[27] and WSe$_2$[39], but only the ground and 2s state have been observed for MoSe$_2$[40]. A closer analysis requires theoretical calculations and is beyond the scope of this paper.

The dependence of the A exciton feature on temperature is shown in 2b. With increasing temperature the distinct absorption at 1.653 eV shifts to lower energies, quenches in intensity and broadens. The shoulder at the lower energy side can be attributed to the reflectance contrast background without absorption (see supplementary S1). At 200K and higher temperatures, this shoulder cannot be as clearly identified anymore owing to the broadening of the excitonic feature in the dielectric function (see supplementary S2). As a result, the error bars increase for the deduced parameters energy, linewidth (FWHM) and amplitude.

The evolutions of energy, linewidth and amplitude with temperature are presented in f. The exciton energy decrease, due to thermal bandgap narrowing, which is in good agreement with the PL temperature dependence[37] and it can be well fitted by the Varshni formula $E_g = E_0 - (\alpha T^2)/(T + \beta)$, where $E_0$ is the energy offset for T = 0K and $\alpha$ and $\beta$ are fitting parameters[41]. The fitting results in $E_0 = 1.653\ eV$, $\alpha = 4.12 * 10^{-4}\ eV/K$ and $\beta$ = 137.7 K, which is in good agreement with previous results[40].

The linewidth follows a steady increase as a function of temperature, typical for phonon-induced broadening. The initial linewidth at 4K (19 meV) is broader than previously observed in PL (12 meV), whereas the linewidth at room temperature (33 meV) is in good agreement with literature PL measurements (34 meV)[25]. However, as the linewidth depends on the substrate and charging condition of the monolayer, different observations are not necessarily in contradiction. Furthermore, the absorption linewidths were compared with PL linewidths that were taken from a smaller illumination area. Averaging over the larger illumination area of about 10 µm, potentially containing more defects or flake edges, can lead to absorption linewidth broadening. The linewidth broadening was fitted by

$$E_x(T) = E_{x,0} + E_{x,B} * T + E_{x,Phonon} \frac{1}{e^{\frac{E_{LO}}{k_B T}} - 1} \qquad (9)$$

Here, $E_{x,0}$ is the exciton linewidth at 0 K, $E_{x,B}$ is the linear broadening constant, $E_{x,Phonon}$ phonon broadening constant and $E_{LO}$ is the phonon energy of the longitudinal optical phonon. $E_{LO}$ was fixed at 30 meV[42], which resulted in fitting parameters of $E_{x,0} = 19.4\ meV$, $E_{x,B} = 1.71 * 10^{-5} \frac{meV}{K}$ and $E_{x,Phonon} = 0.016\ meV$.

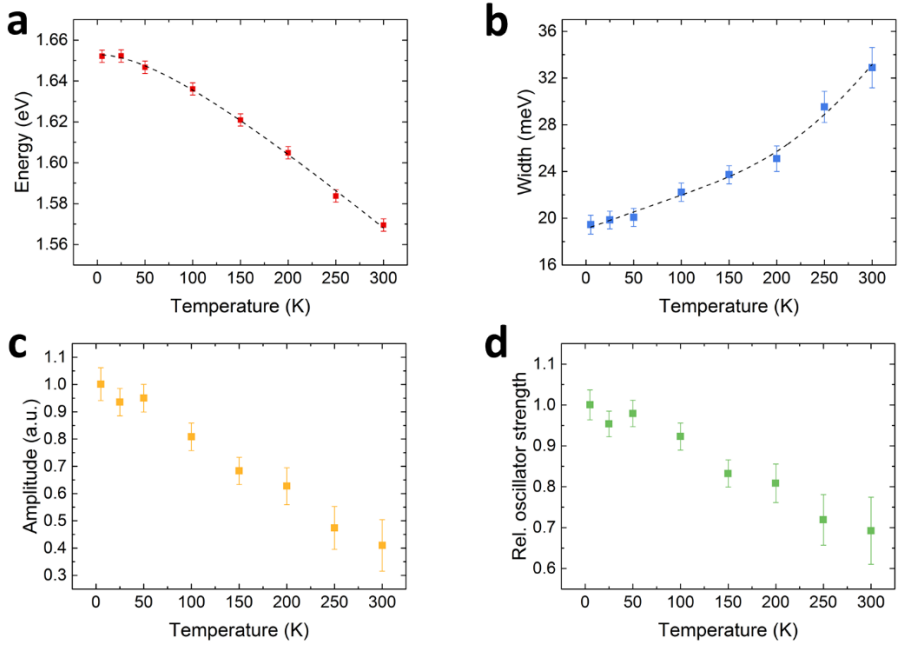

Figure 3: Temperature evolution of the deduced parameters energy (a), linewidth (b) and amplitude (c) and the normalized product of linewidth and amplitude (d), which were used in subsequent calculations.

The amplitude drops almost linearly by 60% from 4K to 300K. This decrease is a natural consequence of the linewidth broadening. However, the product of amplitude and linewidth, a measure for the integrated absorption area, drops also steadily by 30% in the same temperature range, which suggests a temperature-induced drop of the exciton oscillator strength.

The experimentally measured temperature evolution of the exciton linewidth and oscillator strength were used as input for the Rabi splitting and the visibility calculations described in the theory section. The results of both these calculations, coupled oscillator approach (Eq. 1) and numeric TM simulation, are presented in figure 4. The first approach results in low-temperature Rabi splitting value of 17.5 meV. This is in good agreement with the experimentally acquired Rabi splitting of 20 meV[25], which is not surprising as the input coupling strength V was deduced from the experiment. The remaining difference is attributed to the broader exciton linewidth measured in our experiment and the negligence of the lateral mode confinement used in the reference cavity[25]. In the TM simulation, the oscillator strength f (Eq. 4) was adjusted in a way that the simulation result matches 17.5 meV. This procedure provides a realistic estimate for the exciton oscillator strength in monolayer $MoSe_2$, a requirement for the transfer matrix simulations. The Rabi splitting is consistent for both approaches up to 200 K. However, at higher temperatures the results obtained with the two methods diverge with TM simulation results decreasing more rapidly. This slight divergence stems from the simplifications in the coupled oscillator approach. The TM approach is more reliable which is why we exclusively used it for the following calculations. Yet, it should be noted that the calculated reflectivity spectrum for 250 K does not exhibit two clearly distinguishable peaks anymore and the splitting can only be determined by fitting two Gaussian peaks to a broad reflectivity feature. For 300K no splitting can be determined from the simulated spectrum. Both observations are confirmed by the visibility drop below 0.25 for temperatures above 250K. At lower temperature the visibility remains well above 0.25, indicating that the system remains in the strong coupling regime.

The Rabi splitting for the monolithic cavity and for the Tamm plasmon sample is significantly larger (29.3 meV and 33.5 meV at 4K, respectively) and follows a similar decrease as for the open cavity (down to 19.9 meV and 25.0 meV at 300K, respectively). The significant difference compared to the open cavity design is explained by a stronger mode confinement equivalent to a shorter effective cavity length (Eq. 2). Although, the monolithic cavity exhibits a smaller Rabi splitting than the Tamm plasmon sample, the visibilities behave reversely (0.76 and 0.60 at 4K, respectively) due to the narrower monolithic cavity linewidth of 0.2 meV compared to 8.4 meV for the Tamm plasmon structure. However, both visibility evolutions converge towards higher temperatures (0.34 and 0.37 at 300K, respectively). In the Tamm plasmon design, the decrease in oscillator strength does not affect the visibility to the same degree since it is stronger dependent on the cavity linewidth. As a consequence, the calculated visibilities reach the same level at room temperature in both approaches. This level should be high enough to observe strong coupling. Nevertheless, we want to point out that the fabrication of both designs ensuring spectral resonance is more challenging than for the open cavity fabrication. The challenge lies in the overgrowth of the monolayer since conventional deposition methods such as sputtering damage the monolayer. Nevertheless, this task appears achievable since TMDC monolayers have been successfully overgrown by dielectrics[43]. An additional step towards room temperature strong coupling could be the use of multiple, but distinctly separated monolayers as suggested by Dufferwiel et al.[25], which increases the splitting by a factor of $\sqrt{N_{eff}}$ (Eq. 2).

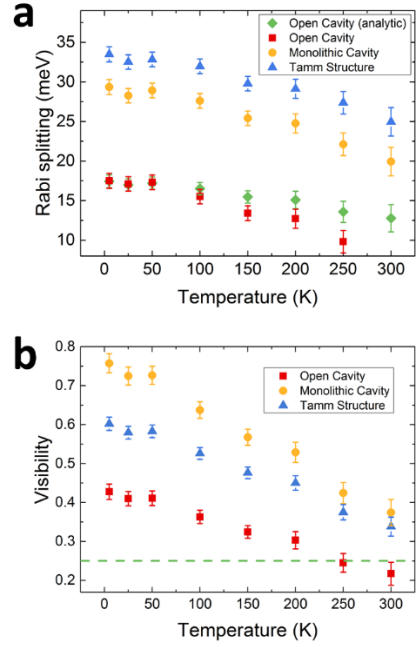

Figure 4: Temperature evolution of the Rabi splitting (a) and the visibility (b) for the open cavitiy design (red squares), the monolithic cavity (yellow dots) and the Tamm plasmon design (blue triangulars). The analytic calculation for the open cavity design was added in a (green diamonds) and the visibility limit of 0.25 was indicated by the green, dashed line in b.

The calculated phase diagram is presented in figure 5. It shows the critical polariton density for different number of MoSe$_2$ monolayers in the system. At T= 1K the density is as low as 3.5*10$^3$ cm$^{-2}$, independent of the ML number. However, at room temperature, it is possible to decrease the critical density from 1.4*10$^{12}$ cm$^{-2}$ to 1.8*10$^{11}$ cm$^{-2}$ by varying from one to ten monolayers due to the increased Rabi splitting, which results in a reduction of the effective polariton mass. Simultaneously, the upper limit (Mott density) rises significantly from 8*10$^{12}$ cm$^{-2}$ for one ML and up to 8*10$^{13}$ cm$^{-2}$ for ten MLs. The comparably high Mott density is due to the small Bohr radius of 2 nm in our system. These calculations assume the following parameters: $m_{ph} = 10^{-5}\, m_e$, $m_{ex} = 0.8\, m_e$[44], $L = 10\, \mu m$, $a_B = 2\, nm$[14]. Rabi splitting for 1 monolayer is taken to be 20 meV. Even more importantly, the upper temperature limit is not defined by the exciton binding energy as for excitons in GaAs (to below 100 K), but only by the strong coupling conditions (the temperature for thermal exciton breaking can be expected to be above the decomposition temperature of the monolayer). We showed that the strong coupling threshold depends on the thermal broadening of the exciton linewidth, the thermal decrease of oscillator strength and the cavity design. Here, we used a visibility value of 0.25 as an indicator for the strong coupling threshold, which yields 250 K for the open cavity design and 400 K (linear extrapolation of the visibility evolution) for both the monolithic cavity and the Tamm plasmon design. For multiple monolayers integrated into any of the structures this limit will further increase as indicated by the shaded area in figure 5. As a result, the phase field for polariton condensation enlarges significantly. Most importantly, the critical polariton condensation density for one monolayer is only $9*10^{11}$ cm$^{-2}$ at 300 K, which is well below the Mott density. For multiple monolayers the range between critical condensation density and Mott density increases even further. From these considerations, it seems feasible that polariton condensation may be observed at room temperature. It should be noted that the achievable polariton density will strongly depend on the polariton dynamics in TMDC monolayers.

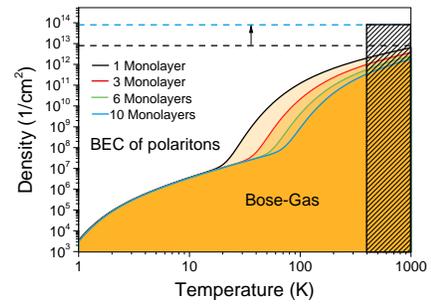

**Figure 5: Phase diagram for a various numbers of MoSe$_2$ monolayers:** Each solid line separates the Bose-gas regime from polariton condensation regime according to Eq. 7. The upper density limit for polariton condensation is given by the Mott density (dashed lines for one (black) and ten monolayers (blue), respectively). The upper temperature limit depends on the strong coupling requirements ( $\vartheta > 0.25$ ) indicated by the shaded area above 400 K (estimated temperature limit for one monolayer).

## Conclusions

We have performed reflectivity measurements on a MoSe$_2$ monolayer at various temperatures between 4K and room temperature. At 4K A- and B-excitons as well as higher excitonic states were identified. The energy spacing between these higher states provides evidence for the non-Rydberg behavior of excitons in MoSe$_2$ monolayers, an effect which is already known from other TMDC monolayers. The temperature dependence of the absorption dip area and linewidth were analysed and used as a basis for subsequent TM simulations of a MoSe$_2$ monolayer integrated into various photonic microstructure designs. As the absorption area, taken as a measure for the oscillator strength, decreases and the linewidth increases, both relevant parameters for strong coupling, Rabi splitting and visibility, decrease with temperature. For the open cavity design this decrease is significant enough to make strong coupling with a single MoSe$_2$ layer hard to be observed at room temperature. In contrast, according to our simulations, strong coupling can be reached even at room temperature for the monolithic and Tamm plasmon structure. A reason for this is the significantly reduced mode penetration into the structures. Finally, we draw the phase diagram for the polariton condensation, which supports the assumption that the condensation of exciton-polaritons may be observed at room temperature in appropriate photonic architectures.

## Acknowledgement

We acknowledge financial support by the state of Bavaria. We acknowledge experimental support by Anne Schade, Isaak Kim and Oliver Iff. AK thanks Mikhail Vasilevskiy for fruitful discussions. EC, AK and SH acknowledge the EPSRC Programme "Hybrid Polaritonics" (EP/M025330/1) for support. ST acknowledges support from NSF DMR-1552220.

Supplementary S1

Energy position, linewidth and amplitude were deduced by fitting a Gaussian function to the reflectance contrast spectra after the background subtraction. In order to carry out an appropriate background subtraction, the reflectance contrast spectra were simulated without excitonic absorptions (imaginary part of the refractive index k of $MoSe_2$ was set to 0). The transfer matrix simulations assume a Si substrate thickness of 200 μm, 285 nm of $SiO_2$ and 0.65 nm $MoSe_2$. Optical constants were either taken from reference[33] or the complex dielectric constants were modelled with a Lorentz oscillator according to equation 4. For comparison a Gaussian shaped dip centered at the resonance was subtracted from the simulated background spectrum. In figure 6, the latter is compared to a simulated background spectrum. The spectra describe the experimentally acquired spectra presented in figure 1b very well. The shoulders above and below the resonance provide an excellent orientation in the background subtraction process. It should be noted that these shoulder are not as distinct in the spectra for 200 K and higher temperatures. Therefore, we assume a higher error in the subtraction and fitting process. The error from the subtraction process was evaluated by conducting the subtraction and fitting multiple times for an identical spectrum. According to this, the energetic position typically varies by +/- 3 meV, the amplitude by +/- 5% and linewidth by +/- 0.75 meV. The total error is composed of the background subtraction error and the fitting error (quadratic error propagation).

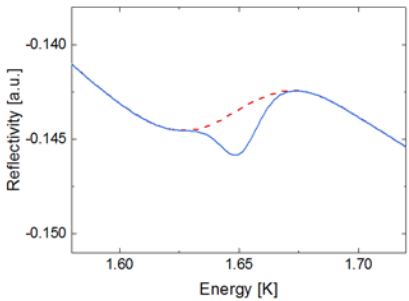

Figure 6: Transfer matrix simulation of the reflectance contrast background without absorption (k = 0) (red, dashed line) and the identical background subtracted by a Gaussian absorption dip at the resonant position (blue, solid line).

Supplementary S2

Figure 7 presents real and imaginary part of the refractive index (n and k) deduced from the Lorenz oscillator model (Eq. 4) and experimental results of linewidth and relative oscillator strength.

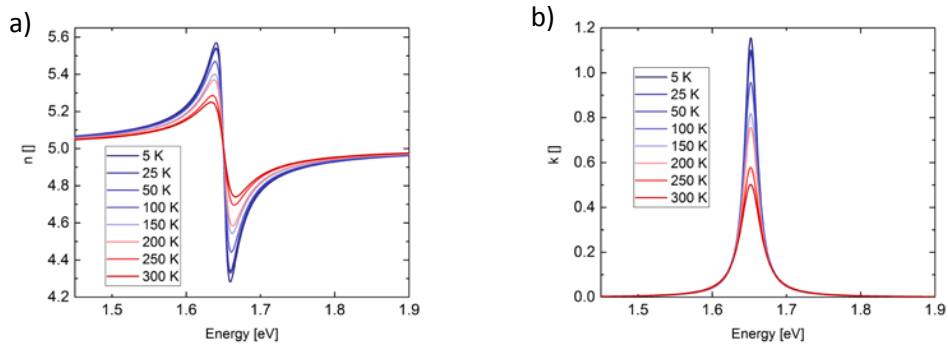

Figure 7: Real (a) and imaginary (b) part of the modelled refractive index of monolayer MoSe$_2$ (for clarity, the spectra are centered at the same energy).

It should be noted that the integral $\int_0^\infty k(\omega, T) d\omega$ follows exactly the same temperature evolution as $f(T)$, which justifies to deduce $f$ from the product of dip amplitude and linewidth, which is in turn proportional to the integrated dip area.

Supplementary S3

The chemical vapor transport technique is used to grow 2H-MoSe$_2$ crystals, where the transport agent is I2. Molybdenum wire (99.95%, Alfa Aesar), Selenium shots (99.999+%, Alfa Aesar) and I2 are sealed in a quartz tube with a vacuum level about 5E-5 Torr. The tube loaded with precursors is placed in a 3-zone horizontal furnace. Two ends of the tube are kept at 1085 °C and 1030 °C separately, causing a temperature deference of 55 °C [1].

[1] Wildervanck, J. C. Chalcogenides of Molybdenum, Tungsten, Technetium and

Rhenium. PhD thesis, Univ. of Groningen (1970)